# GENDERING OF SMARTPHONE OWNERSHIP AND AUTONOMY AMONG YOUTH: NARRATIVES FROM RURAL INDIA

Renza Iqbal, Erasmus University Rotterdam, renza@esphil.eur.nl

**Abstract:** This study delves into the research question: how does gender influence smartphone ownership and autonomy in using the internet among the youth in rural India? This paper explores the influence of local culture on smartphone ownership and autonomy through an ethnographic study among rural Indian youth by analysing the intersection of gender with other identity axes. The findings show that young people's smartphone ownership and autonomy is shaped by their social and cultural setting, and could lead to various inequalities in their internet usage. This study shows that gender paves way for various disparities with regard to smartphone ownership and internet usage. Decolonisation of the understanding of smartphone ownership and internet usage patterns of the youth in the Global South suggests a reconsideration of the user experience designs and platform policies.

**Keywords:** smartphone, autonomy, privacy, ICT, Global South, gender and ICT

## 1. INTRODUCTION

In 2016, the UN declared access to the internet as a basic human right (Sandle, 2016). GSM Association (2020) estimates that 600 million new subscribers will access the internet globally through mobile phones by 2025. In most low-and-middle-income countries, a smartphone is often the only device affordable to a vast majority to access the internet. The launch of cheap data plans by Reliance Jio Services in 2015 led to an increased number of mobile phone users in India (Sengupta, 2017). This paper argues that access to devices or the internet alone does not meet the needs of a diverse range of users across the globe. Several cultural underpinnings are involved in deciding the availability and accessibility of the internet on smartphones.

This study is informed by ethnographic fieldwork undertaken in rural Kerala, a state in the South of India, in 2019. This paper explores the understanding of smartphone ownership and internet usage among young people and how their understanding and notions of autonomy influence their internet usage and online behaviour. Across most communities in India, senior family members keep a constant watch over their womenfolk for what is perceived to be in the interest of their safety. Localized gendered narratives of autonomy significantly impact the degree of access available to young women in rural India.

This study on young users in rural India who access and consume the internet through their smartphone reveals that autonomy is defined variedly across communities and is deeply gendered in everyday accessibility and usage. It is essential to analyze the different factors that influence a community's culture of usage to draw connections between traditions, norms, and understanding of digitality. Cultural factors such as rurality, gender, religion, and class play an important role in influencing such habits and aspirations. Studies have shown that cultural norms control and limit women's use of technology in South Asia (Sambasivan et al., 2018). This paper calls for the need in understanding localized narratives of autonomy catering to the needs of the individual user.

## 2. LITERATURE REVIEW

The major obstacles to internet adoption in South Asia are a large gender gap and a lack of adequate skills (GSM Association, 2020). Compared to men, women in low-and-middle-income countries are 23% less likely to use mobile internet; in South Asia, the mobile internet use of those living in rural areas is 45% less than those in urban areas (Rowntree et al., 2020). Many communities in the Global South impose strict regulations restricting women's access to public spaces in the interest of their safety. These restrictions are now transcending the physical social space they occupied and slowly moving into the digital space. There is a systemic denial of digital access to women in India, which involves both economic and normative barriers (Barboni et al., 2018).

Most rural digital exclusion studies point to the lack of motivation and skills as a hindrance to digital engagement (Salemink et al., 2017). Digital inclusion is to be comprehended as more than access and look into class formation, stratification of information, structural systems of power, local cultures, and their influence on the user's autonomy. Digital inclusion does not just pertain to the have's and have nots; the uneven access reveals the existing social stratification. The global privacy narrative focuses on concerns regarding data collection, data control and third party sharing (Mahmoodi et al., 2018; Van Dijk et al., 2018) and fails to acknowledge local privacy narratives influenced by culture.

In many parts of the Global South, cultural expectations dictate that women should share their mobile phones with family members and that their digital activities be open to scrutiny by family members (Sambasivan et al., 2018). Therefore, understanding the cultural substructures that lead to this situation is imperative in comprehending these obstacles. This paper draws attention to how localized notions of autonomy lead to the differences in users' privacy, skills and ability to access, thereby demanding recognition of the gendered definitions of autonomy in underrepresented contexts.

## 3. METHODOLOGY

The study seeks to answer the research question: how does gender influence smartphone ownership and autonomy in using the internet among the youth in rural India? using ethnographic methods. Ethnography creates contexts through which the researcher delves into the participants' lives using interventional and observational methods to reveal what matters to the participants (Pink & Morgan, 2013). In-depth interviews in the native language help capture individual experiences and cultural practices (Schensul et al., 1999). The researcher engaged first-hand with the participants in a deliberate attempt to understand their lives and experiences. For the purpose of this study, youth has been defined as those who are between 15 to 25 years of age. This study looks into the nuanced negotiation between smartphones and gender.

Observation and in-depth interviews were engaged in fields across nine Panchayaths in the rural areas of Wayanad district, in the north of Kerala, India, during June 2019. The study consisted of a focus group discussion (FGD) and twenty-five in-depth interviews conducted in person by the researcher. The respondents were chosen through purposive sampling to include diverse age, gender, religion, caste, educational qualification, and economic condition. The respondents included students, working people, and homemakers. The income groups include respondents belonging to the low to upper-middle-class; the religious backgrounds included Hinduism, Islam, and Christianity; and both male and female participants. The respondents were identified with the help of grassroots level ASHA (Accredited Social Health Activist) workers from various localities across Wayanad. The ASHA workers form a close network and maintain a strong bond with all the families in their purview. They carry the most updated demographic information regarding the households in their region. Approaching households via ASHA workers and being accompanied by a local helped gain the trust of the respondents and their families.

The data collection attempted to understand their smartphone ownership and autonomy: the devices used, ownership of the device, the autonomy exercised, skills acquired, restrictions associated with usage imposed by kin, and the frequency of use, among others. Focus group research helps identify pertinent issues, areas for further investigation, meanings, values, opinions, behaviours, and explanations for cultural or physical phenomena (Schensul et al., 1999). Students from the Calicut University Centre at Wayanad were invited to participate in the focus group discussion. An equal representation of male and female students was ensured. The participants were from four different Panchayats and had diverse religious and economic backgrounds. Permission for the FGD was obtained from the authorities at the Centre. The Centre chose students from those who were interested to participate after being informed about the purpose of the FGD. The group consisted of eight members who were familiar with each other and shared a level of comfort. Consent was obtained from the participants, and the confidentiality of the participants has been safeguarded. Names of the participants have been changed to protect identities.

Face to face, semi-structured in-depth interviews were conducted at the respondents' residence in their local language to make the respondents comfortable. Locals accompanied the interviewer during the process to validate the authenticity of the interviewer to the respondents and help build a rapport. Being introduced by locals helped build trust, ensuring genuine research participation. The interviews and the FGD were audio recorded with the respondents' oral consent, and the recordings were later transcribed. This was followed by thematic coding and analysis. The coding involved: development of a coding scheme, applying the coding scheme, ensuring consistency of its application (Schensul et al., 1999). The themes were identified from a combined understanding of the data and literature.

## 4. FINDINGS AND ANALYSIS

A smartphone is the most commonly used device to access the internet, affordability being the prime reason. Of the ten people that did not have a personal smartphone, seven were female. It was also observed that most of the respondent's fathers had smartphones, but not all mothers did. There is a gendered difference in sharing devices, autonomy over devices, leisure time, and conduct in public spaces. An interplay of social and cultural factors constantly influences rural women's narrative and definition of autonomy over their devices. They perceive and interact with multiple meanings of autonomy and differentiate it as essential or inessential and determine its significance in their day to day lives.

The respondent's narratives on their ownership and use of the internet on smartphones reflected a stark gender difference. The findings were organized based on the recurring themes identified from the narratives. The various gendered themes brought out were: age of ownership – male users owned a smartphone significantly earlier than female users, code of conduct – there were expectations from female users regarding how to or not to use the internet through devices, lack of autonomy over personal devices – female users' devices were expected to be shared or monitored, social media usage was guided by societal expectations, nothing to hide – young women's internet usage was monitored in the interest of their safety, and therefore they were expected to cooperate with the same, repercussions in the event of deviance – if the woman breaks the prescribed code of conduct her smartphone was withheld from her. Women in the community are expected to follow a code of conduct in public places. The study shows that these regulations transcended into their private space of internet usage and mobile phone accessibility. The study reveals that in predominantly patriarchal rural Indian societies, young women are deprived of autonomy in internet usage over safety concerns. Autonomy and privacy thus took a different meaning for young women in rural India while navigating through these cultural contexts.

### 4.1. First time ownership

"Boys get smartphones much before girls," Hasna (20, f) was visibly upset with the imbalance. In 96% of the households, the male child got a personal smartphone much earlier than the female child. The average age at which a male user first gets ownership of a smartphone is 15, while it is 18 for female users. Habeeb (21, m) explains that "parents are apprehensive about giving a phone to a girl; they are sceptical about their security. Boys grow up in a freedom-oriented atmosphere." "When we ask our parents for a phone, they ask us 'what is the need?' proposing that 'if you need to call friends there is a phone in the house," shared Ramitha (21, f). When boys are free to go outside, do things they like, girls are expected to explain and justify their need to go out.

## 4.2      Financial independence & ownership

"Many of my male classmates made money and bought a phone for themselves," shares Ansiya (18, f). Boys tend to seek part-time jobs and other forms of employment during the summer school break, making enough money to buy themselves a phone, whereas girls are not free to do the same owing to cultural barriers. For young women from well to do families, engaging in employment is seen as a disgrace on the family, as the incapacity of the primary breadwinner to provide sufficiently; therefore, they have to rely on the decision-makers in the households to receive a personal smartphone. Rural areas suffer from conservative social contexts that influence digital disengagement (Sora & Kim, 2015; Salemink et al., 2017). Women carrying out paid labour was understood to harm their domestic work and to the essence of femininity - along with which there is a fear of being exposed to sexual harassment when working outside, and women who challenge these norms are often ostracized  (Devika, 2019).

However, it was observed that it was acceptable for women belonging to economically backward tribal communities to engage in employment. However, those among them who do not have the opportunity to work often never have a phone for themselves, owing to financial constraints. Surya (20, f) is a young mother belonging to Paniya, a tribal community in Wayanad. She does not generate any income, own a personal device, or engage with the internet. Her only engagement with the internet is, "when he (her husband) shows me things, I look at it." Shidusha (22, f) and Sruthi (22, f) belong to Kuruma, a tribal community in Wayanad. While the former bought a phone for herself when she started working as a teacher, the latter used her scholarship funds to get a personal smartphone. Nimisha (26, m) – another respondent from the Kuruma community, has been unable to go for a job as she has been taking care of her toddler and does not have a personal phone.

## 4.3      Leisure

Young women have less leisure time than their male counterparts, which translates to comparatively less time spent on the internet. "Girls don't get to spend much time on the internet. They come to college in the morning, go back home and study, engage in household chores. Unlike us, they are scolded by family if they overspend time on the phone," says Arunjith (18, m). Patriarchy constrains women's leisure through social structures (Shaw, 1994). Female respondents stated that the internet is only to be used in a time of need, as opposed to the male respondents who relied on it primarily for leisure. Abhijith (22, m) loves the internet, "I have two sim cards and consume 2.5 GB a day, which often falls short. I recharged again on my way home now." Arunjith (18, m) claims that he is on his phone all the time and replies instantly to all messages. With male respondents spending significantly more time on various platforms through their devices, it is not surprising that young men are more skilled at the nuances of using a smartphone.

The cultural expectation defines that a good woman does not engage in a lot of leisure, as she is busy with her chores; thus women often feel that they are not entitled to leisure (Harrington et al., 1992; Shaw, 1994). Young female respondents attempted to sound compliant to the local moral codes associated with digital usage. Dheeraj (17, m) does not shy away from admitting that he uses the internet whenever possible, playing video game's such as Player Unknown's Battlegrounds (PUBG) and being active on Facebook, TikTok, and Instagram. He shared an incident from their neighbourhood, where a young woman fled home after being restricted from using Facebook as she pleased. For female users, their right to leisure is also restricted, as is their right to autonomy.

## 4.4 Likelihood of Intervention

Rama's (23, f) family knows the password to her smartphone, and she thinks it is okay if they are observing and enquiring about her activities in a friendly manner. When asked if a personal smartphone increases one's privacy, Jancy (23, f) says, "we (Jancy and her brother) share everything. I have not felt a need for privacy." However, the sharing does not involve Jancy looking into her brother's smartphone. All respondents agree that if the user is careful, they will be free from trouble, subtly pointing at the pertinent victim-blaming culture.

Most of the participants likened the perspective that autonomy over the smartphone was a privilege reserved for men. Keeping women's devices open to be used by family members and having nothing to hide was considered culturally appropriate. "Unlike girls, we use our phones all the time. We use apps and other techniques to protect our device from being hacked. They (girls) would know little about the technicalities," Arunjith (18, m) explains why female users are more at risk. The primary concerns and risks around a female user's online safety are centred on the probability of them making uninformed decisions while interacting with strangers, being unable to protect the information on their devices, leading to a misuse of the same, and engaging in romantic affairs.

## 4.5 Autonomy

Nimisha's (25, f) mother often watches over what she is doing on her mobile phone and asks whom she is talking to or chatting with. Rama's (23, f) parents question her when she spends long hours on her smartphone. Asfeena's (25, f) mother-in-law took a seat with us to listen to the entire interview where Asfeena told us that she does not have a smartphone. She briefly mentioned that she occasionally watches videos on YouTube on her husband's phone but then said that she does not use the smartphone. Fasna (19, f) got a smartphone after her marriage. She is happy that now she can use WhatsApp freely; earlier, she used it on her father's phone, "my husband occasionally scrolls through my phone, not that he is checking or anything." The monitoring of women's smartphones by family members is dictated by culture, and they are expected to keep their digital activities open to scrutiny (Sambasivan et al., 2018).

None of the male respondents was subjected to such monitoring; they enjoyed absolute autonomy over their internet usage and smartphones. "Parents hear about various issues arising from the use of the smartphone at awareness sessions and thus develop a fear," says Shruti (22, f). The usage of female respondents was being monitored regardless of their age or marital status. "Boys' phones have locks, on top of that a fingerprint lock, over that a pattern lock," says Amal (22, m). Female users are not free to secure their phones or apps as they are expected to have nothing to hide. He suggests, "parents don't give phones to their daughters because then they will have more freedom which could get them into trouble. Boys can save themselves from any problem." The nothing to hide argument is subtle oppression in disguise (Devika, 2019).

Abhijith (22, m) shared insights on how male users often get into trouble by befriending women online. "At first, she will ask for a mobile phone recharge. Stupid boys would do it for them. This would keep repeating. Then these girls will go out with these boys. Then the boys end up spending over that – at movies, coffee shops or otherwise. I have personal experience, and it is not just me. I know many who have been trapped similarly. The girl ghosts later," however such incidents do not affect their autonomy over their devices.

Jancy's (23, f) family told her not to talk on the phone when outside and to restrict phone conversations to when she is within the house. Otherwise, people in the locality would assume that she is in a relationship. She shares that female friends whose families did not provide them with a smartphone were sometimes given one by their boyfriends, which they used without the knowledge of their family. Strict cultural norms limit the public spaces to men, and women are often expected to recede indoors.

### 4.6 Honour

Traditional gatekeepers of social morality keep a check on the social customs and conventions, regulating what is allowed and for whom. However, these restrictions are bound only to female users, and male users act as gatekeepers. These gatekeepers could take away the autonomy of a woman who breaks social norms. When young people in these communities engage in romance, it is most often in secret as it is not socially acceptable. Hasna's (20, f) smartphone was confiscated by her family when her romance with a fellow college-mate was 'caught'. Local observers fear that women occupying public spaces would engage in sexual or romantic relationships outside marriage (Kaya, 2009), which continues to be one of the prime fear factors concerning women's digital access and usage, especially in inherently patriarchal communities. Romance itself is belittling in these communities, and elopement would put the girl's family to shame. The family withholds women's autonomy as women are destined to uphold the family's virtue (Arora, 2019).

These norms do not apply to male users. "I spend over four hours on my smartphone," shares Abhijith (22, m), whose internet usage is seldom interfered with and never monitored. "Only half of my female classmates have a phone. They are often not given one by their families as there is a concern about them making new friends over Facebook," says Amal (18, m). Amal (18, m) shares that "a girl from my school was caught bunking class and hanging out with some friends she made online. The teachers and the parents believe that it was her phone that led her 'astray'. Post this, girls at our school are now denied access to phones. School authorities and their parents confiscated their phones". An authoritarian society adopts a paternalistic role when it comes to the autonomy of women and girls if it identifies autonomy to be at odds with the social harmony of the society. "They (parents) wouldn't want their daughters to be trapped in love. They might refuse to get a phone for their daughters to protect them," opines Arunjith (18, m). Social norms can overtake consent where privacy and morality is dominantly a gendered space (Arora, 2018).

### 4.7 Social media

Social media use of female users is highly regulated and closely observed by gatekeepers in the community. There is a constant fear among young women about their photos on social media being used in fake pornographic content or personal messages being publicized (Arora 2019). The female users carry the burden of keeping themselves safe from the troubles online. Jancy's (23, f) brother warned her against the possible misuse of photos in one's phone if taken in compromising positions such as nudes or intimate images - if the phone needs to be repaired, the person repairing it can get hold of such pictures. Rama (23, f) was instructed by her elder brother not to post any of her personal pictures on Facebook. Such restrictions often lead to the abstinence of women from the digital worlds.

A recurrent comment from the female respondents was, "if we are careful, everything will be fine," which shifts the focus from the perpetrator to the victim. The most commonly faced issue on social media was of being approached by strangers seeking friendship which they then attempt to convert into a romantic relationship. Women who decline bear the brunt of offensive messages, comments, or the threat of having their number shared with several others who constantly send disturbing messages. Shilpa (19, f) received several inappropriate messages from strangers when she was on Facebook – she would immediately block them on the platform; she quit Facebook after getting married. Facebook has a poor reputation in the community.

## 5. DISCUSSION

A dominant narrative identified during the fieldwork is that female users were less concerned about the popular global privacy narrative regarding data being scooped by companies or Governments. They claim that autonomy over their smartphone was not necessary for them and the constant monitoring of their mobile internet usage was in their best interest. Most young women approve of this action and therefore are less assertive of a need for autonomy while using a personal device. The study suggested that this void of autonomy and monitoring for safekeeping were hardly applicable to men. While women claimed that they had nothing to hide, male respondents claimed to be adept at protecting their data from the surveillance of their family members. Young women in rural India are claiming that they do not need autonomy, presenting an alternative narrative to that of their counterparts in the West, who may take autonomy in using and accessing the internet through a personal device with non-interference from family members for granted.

However, during the focus group discussion, in the absence of their family, the female respondents shared in unison that this gendered monitoring is unfair and that "we know how to overcome it as we use the phone more than our parents." In the presence of kin, female respondents sounded indifferent to the need for autonomy over their devices. They tried to portray this indifference as a virtue, while male respondents proudly proclaimed how they use various security mechanisms to keep everyone else out of their devices. "All apps on my phone are locked and shall unlock only on detecting my face," Abhijith (22, m). User privacy for smartphones is designed based on the understanding that one device shall be used by one user. However, the lack of autonomy of the users challenges this definition and seeks to rethink the available privacy control options.

## 6. CONCLUSION

The study shows that the notions of autonomy in using smartphones are vastly gendered, both in terms of meaning and access. It was evident that the skillset, peer support, and freedom to use the internet on smartphones vary remarkably based on gender. The responses from the participants indicated that no female respondent had absolute autonomy over their smartphone; their devices were either shared with other family members or closely monitored by their kin. This owes to cultural restrictions and the hierarchies existing in the society, which puts male members before the female members - owing to the inherently patriarchal nature of these societies - as seen in the rural areas of developing countries in the Global South. Decolonising the inherent understanding of privacy and autonomy is essential in ensuring that everyone gets equal access and opportunity in the digital space. The study shows that penetration of the ICT's alone does not ensure equal participation. The findings emphasise the need for developing design and policy with the local culture and local user's needs in mind, thus deviating from making universal design/policies based on the preferences, requirements and needs of the Global North. The findings of this study are limited to users' perceptions in rural Wayanad and encourage future research to replicate this study in diverse contexts by including the perspectives of other stakeholders such as designers and policymakers.

## REFERENCES

Arora, P. (2018). Decolonizing Privacy Studies. *Television & New Media*, 20(4), 366–378. https://doi.org/10.1177/1527476418806092

Arora, P. (2019). General Data Protection Regulation—A Global Standard? Privacy Futures, Digital Activism, and Surveillance Cultures in the Global South. Surveillance & Society, 17(5), 717–725. https://doi.org/10.24908/ss.v17i5.13307

Barboni, G., Field, E., Pande, R., Rigol, N., Schaner, S., & Moore, C. T. (2018). A Tough Call: Understanding barriers to and impacts of women's mobile phone adoption in India. Harvard Kennedy School-Evidence for Policy Design.

Devika, J. (2019). Women's Labour, Patriarchy and Feminism in Twenty-first Century Kerala: Reflections on the Glocal Present. Review of Development and Change, 24(1), 79-99.

GSM Association. (2020). The mobile economy–2020. GSMA Intelligence, 30. Retrieved from https://www.gsma.com/mobileeconomy/#key_stats

Harrington, M., Dawson, D., & Bolla, P. (1992). Objective and subjective constraints on women's enjoyment of leisure. Loisir et société/Society and Leisure, 15(1), 203-221.

Kaya, L. P. (2009). Dating in a sexually segregated society: Embodied practices of online romance in Irbid, Jordan. Anthropological Quarterly, 82(1), 251-278.

Mahmoodi, J., Čurdová, J., Henking, C., Kunz, M., Matić, K., Mohr, P., & Vovko, M. (2018). Internet users' valuation of enhanced data protection on social media: Which aspects of privacy are worth the most?. Frontiers in psychology, 9, 1516.

Pink, S., & Morgan, J. (2013). Short-term ethnography: Intense routes to knowing. Symbolic Interaction, 36(3), 351-361.

Rowntree, O., Shanahan, M., Bahla, K., Butler, C., Lindsey, D., & Sibthorpe, C. (2020). The Mobile Gender Gap Report 2020: GSMA. Retrieved from gsma.com/r/gender-gap/

Salemink, K., Strijker, D., & Bosworth, G. (2017). Rural development in the digital age: A systematic literature review on unequal ICT availability, adoption, and use in rural areas. Journal of Rural Studies, 54, 360-371.

Sambasivan, N., Checkley, G., Batool, A., Ahmed, N., Nemer, D., Gaytán-Lugo, L. S., ... & Churchill, E. (2018). " Privacy is not for me, it's for those rich women": Performative Privacy Practices on Mobile Phones by Women in South Asia. In Fourteenth Symposium on Usable Privacy and Security ({SOUPS} 2018) (pp. 127-142).

Sandle, T. (2016, July 23). UN thinks internet access is a human right. Business Insider. Retrieved from https://www.businessinsider.com

Schensul, J. J., LeCompte, M. D., Nastasi, B. K., & Borgatti, S. P. (1999). Enhanced ethnographic methods: Audiovisual techniques, focused group interviews, and elicitation. Rowman Altamira.

Sengupta, D. (2017, Sep 05). Reliance jio infocomm's launch disrupts telecom landscape [internet]. The Economic Times Retrieved from https://www-proquest-com.eur.idm.oclc.org/newspapers/reliance-jio-infocomms-launch-disrupts-telecom/docview/1935039538/se-2?accountid=13598

Shaw, S. M. (1994). Gender, leisure, and constraint: Towards a framework for the analysis of women's leisure. Journal of leisure research, 26(1), 8-22.

Sora, P., & Kim, G. (2015). Same access, different uses, and the persistent digital divide between urban and rural users. In The 43rd Research Conference on Communications, Information and Internet Policy (pp. 1-23). SSRN.

Van Dijk, N., Tanas, A., Rommetveit, K., & Raab, C. (2018). Right engineering? The redesign of privacy and personal data protection. International review of law, computers & technology, 32(2-3), 230-256.